%% file: blrms_submission.tex
\documentclass[12pt]{iopart}
\usepackage[dvips]{graphicx}
\usepackage{color}
\usepackage{url}
\usepackage[dvips]{hyperref}

\begin{document}

\title[LIGO seismic noise long term study]
{Long term study of the seismic environment at LIGO}
\author{E J Daw \dag \ddag \footnote[3]{To whom correspondence should be 
addressed}, J A Giaime \dag , 
D Lormand $\|$, M Lubinski \P\ and J Zweizig $^+$}

\address{\dag\ Department of Physics and Astronomy, Louisiana State University,
Nicholson Hall, Tower Drive, Baton Rouge, Louisiana 70803-4001, U.S.A.}

\address{\ddag\ University of Sheffield, 
Department of Physics and Astronomy, Hicks Building, Hounsfield Road,
Sheffield S3 7RH, England, U.K.}

\address{$\|$ LIGO Livingston Observatory, 19100 Ligo Lane, Livingston, 
Louisiana, 70754, U.S.A.}

\address{\P\ LIGO Hanford Observatory, Route 10, Mile Marker 2, Richland,
Washington 99352-0159, U.S.A.} 

\address{$^+$ LIGO Laboratory, California Institute of Technology,
Pasadena, CA 91125, U.S.A.}

\ead{e.daw@shef.ac.uk}

\begin{abstract}
The LIGO experiment aims to detect and study gravitational
waves using ground based laser interferometry. A critical factor
to the performance of the interferometers, and a major consideration
in the design of possible future upgrades, is isolation of the 
interferometer optics from seismic noise. We present the results of
a detailed program of measurements of the seismic environment
surrounding the LIGO interferometers. We describe
the experimental configuration used to collect the data, which
was acquired over a 613 day period. The measurements focused on the frequency
range 0.1--10~Hz, in which the secondary microseismic peak and noise
due to human activity in the vicinity of the detectors was found
to be particularly critical to interferometer performance. 
We compare the statistical 
distribution of the data sets from the two interferometer sites,
construct amplitude spectral densities of seismic noise
amplitude fluctuations with periods of up to 3 months, and 
analyze the data for any long term trends in the amplitude of seismic
noise in this critical frequency range.
\end{abstract}

\submitto{\CQG}

\maketitle

\section{Introduction}
\label{sec:introduction}

The LIGO laboratory has constructed three long-baseline laser interferometers
whose object is the direct detection and subsequent study of gravitational 
radiation from astrophysical sources \cite{abramovici}. 

\begin{figure}
\begin{center}
\input{bothsites_horiz.pstex_t}
\end{center}
\caption{\label{fig:schematic}
A simplified schematic diagram showing the layout of the LIGO
interferometers, and the 
approximate positions of the seismometers at the sites. Note
that the cartesian axes define the three components of velocity relative
to the interferometer arms. The direction of true north with respect to the
interferometer arms for each site is indicated by the arrow labelled with N.
By true north, we mean the direction parallel to the line of constant longitude
passing through the corner station building of the LIGO site in question.}
\end{figure}
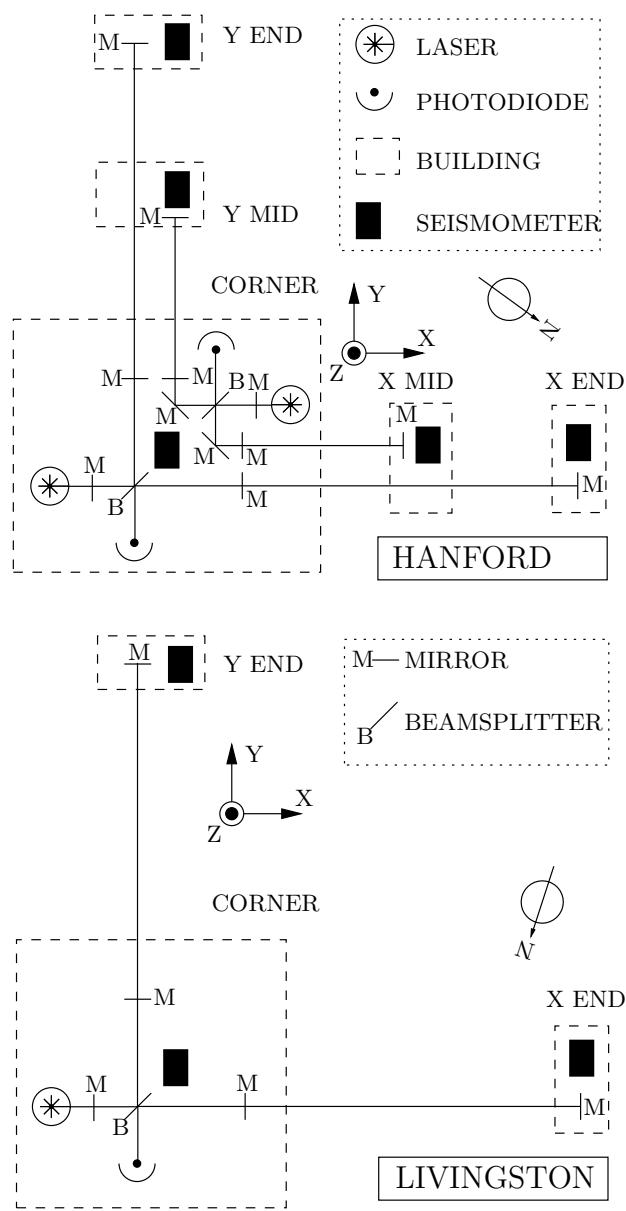

The LIGO interferometers are located at two sites in the U.S.A., 
two interferometers of arm lengths 2km and 4km
at the Department of Energy Hanford Site near Hanford, 
Washington, and a single 4km arm length interferometer 
located in Livingston Parish, Louisiana.
We refer to these two observatories subsequently as Hanford 
and Livingston. Figure \ref{fig:schematic} shows greatly simplified schematics of the
interferometers.

The LIGO detectors have been described in detail elsewhere \cite{instpaper}.
Each detector consists of a Michelson interferometer
with the two interferometer arms at right angles. Each arm consists of a
2 or 4 kilometer long Fabry-Perot resonant cavity. A laser beam is split 
into two components which coherently excite 
one resonant mode of each Fabry-Perot
cavity. At resonance, 
where the cavity length $L$ is a multiple of half the laser wavelength $\lambda$,
there is no phase shift between the beam incident 
on the near-end mirror of each 
cavity and the beam passing back towards the beam splitter 
from this same near-end mirror. For small deviations, $\Delta L$,
from the resonant length, the phase shift is linear in $\Delta L$, 
as long as $\Delta L$ is small
compared with $\mathrm{\Gamma_L = \lambda/2} \mathcal{F}$, 
where $\mathcal{F}$ is
the finesse of the cavity. The LIGO interferometers have
$\mathrm{\lambda = 1.06~\mu m}$ and $\mathcal{F} \sim \mathrm{100}$, so
$\mathrm{\Gamma_L \sim 5~nm}$.
The detectors are designed to be sensitive to gravitational waves in 
the frequency range 50~Hz to 7~kHz. 

Gravitational waves passing through the detectors would cause 
oscillatory perturbations in components of the metric. Consider 
first a simplified case, where the mirrors are freely floating, initially
at fixed separation. 
Metric perturbations due to gravitational waves would cause 
the lengths of the two Fabry-Perot 
cavities in each interferometer to oscillate in antiphase. 
The amplitude of these oscillations from anticipated sources of 
gravitational waves are not known, but they are not
expected to exceed $\Delta L/L = 10^{-21}$. These oscillations 
would give rise to phase modulations of opposite signs in the light beams 
reflected from the near end of the two Fabry-Perot cavities back towards 
the beam splitter. By recombining these beams coherently at the beam
splitter, we can measure the phase difference between them.

The response of the LIGO interferometers to gravitational waves is
more complex than for the ideal case discussed above.
Each mirror is supported against the Earth's gravitational field. The reaction 
forces in this support system must be taken into account in 
calibrating the response of the LIGO interferometers to gravitational waves.
Secondly, the LIGO interferometers utilize multiple feedback \cite{lsc}
and feedforward \cite{msff}
control systems to minimize residual fluctuations in the 
lengths of the Fabry-Perot cavities.
These control systems utilize electromagnetic actuators
to exert forces on the LIGO mirrors to ensure that fluctuations
in the arm lengths over the full interferometer bandwidth
are sufficiently small to maintain linearity of the interferometer and
readout electronics. The reaction of these control systems to gravitational
waves must also be accounted for in calibration of the interferometers
\cite{lsc, gaby}. 

There are more stringent constraints on the magnitude of interferometer
arm length fluctuations than the $\mathrm{\pm 5~nm}$
required to ensure linearity in the response of the phase of the recombined signal at the beam
splitter to length fluctuations. The interferometer optics are configured for a
light intensity at the output photodetectors commensurate with high 
sensitivity to the small arm length fluctuations expected from 
gravitational waves. In this configuration, the photodetectors
saturate for arm length fluctuations exceeding $\mathrm{1~pm}$. 
Furthermore, length fluctuations approaching this photodiode limit can
result in excess noise in the gravitational wave band through bilinear
coupling of laser amplitude and frequency noise into the interferometer. 
Fluctuations in the arm lengths are maintained to within these stringent limits
by the servo controls described above.

A problem of great practical importance is {\it lock acquisition}, the
process of bringing the mirrors from initially unconstrained motion on
their supports, into a state where their residual motions
are sufficiently small that the interferometer outputs are linear in 
changes in the arm lengths due to mirror displacements, 
and control systems are operating to 
maintain the resonant, high sensitivity state of the instrument
\cite{lockacquisition}. We refer to loss of this sensitive instrument
state as {\it lock loss}. When lock loss occurs, the interferometers are
unuseable as gravitiational wave detectors until lock is re-acquired.
The operation of the LIGO interferometers is largely concerned
with maintaining lock for the largest possible percentage of the time.

A significant noise source for LIGO is seismic noise in the
ground acting through the mirror supports and shaking the mirrors.
In order to be sensitive to gravitational waves, the amplitude
spectral density of noise fluctuations on the mirrors due to seismic
noise over the 50--7000~Hz band should be significantly less than the
expected signal amplitude of $\mathrm{\sim 10^{-18}~m}$. Secondly, the 
root mean square (r.m.s.)\ noise
amplitude at each mirror over all frequencies should be significantly 
less than the $\mathrm{10^{-12}}$~m required to maintain lock.

As part of the LIGO scheme to meet these displacement noise requirements, each
mirror support incorporates a multi-stage vibration isolation system.
This system consists of a passive isolation stack \cite{passivestack}
in series with a steel wire, single loop pendulum 
for which the mirror substrate forms the suspended mass. Both of these
systems have natural resonant frequencies below the gravitational wave
detection band. The 0.74~Hz pendulum resonance is damped using a local
feedback control system. The passive isolation stack has
several resonances between 1 and 15Hz, characterized by Qs of about
10--30. 

In this paper we present results of a long term study of seismic noise
at the two LIGO sites. 
The seismometer data recorded for this study reflect
motion of the concrete slabs under the LIGO optic support structures. The
results given here are not intended as a study of the seismic environment
in the ground. To reflect this, we use the terms slab velocity and slab
motion throughout this paper to denote the measured ground velocity on 
the concrete floors of the LIGO lab buildings, as opposed to measurements
of ground velocity in a seismic vault that would be used by a geologist or
geophysicist to study the geophysical properties of the local earth's crust.

The data for this analysis were taken over a
613 day period between April 2nd 2001 and December 10th 2002. The analysis
focuses on the frequency band 0.1--10~Hz. A central goal of this type of
analysis is to study periodicities in the seismic environment at the sites and
any correlations they might have with performance of the interferometers.
It became clear during this
analysis that slab noise in different parts of this frequency band was
a good predictor of interferometer functionality. Furthermore, this analysis
has demonstrated conclusively that the seismic environment at the
Livingston interferometer is more challenging to ground-based 
gravitational wave interferometry than that at the Hanford site. 

\section{Locations and Environments of the LIGO Detectors}
\label{sec:location}

Herein follows a brief summary of the location of each of the two
LIGO interferometer sites and environmental information 
about the two sites which is relevant to our analysis of the seismic
data.

The Hanford interferometers are located on semi-arid shrub and
grassland east of the Cascade mountains in Washington State
on the Columbia Plateau, 370~km 
from the nearest point on the Pacific coast.
The vertex latitude and longitude of the Hanford laboratory are
$46^{\circ}27'18.5''$ N, $119^{\circ}24'27.6''$ W, and the x arm
orientation is $324.0^{\circ}$ (NW).
The plateau rests on a sequence of volcanic basalt flows called
the Columbia River Basalt Group. These basalt formations are folded
and faulted, and the resulting basins have filled with alluvial sedimentary deposits.
The Hanford interferometers rest on the sediment of one of the
larger basins known as the Pasco basin. The sediment depth exceeds the
depth of the water table, which is approximately 120~m below the surface \cite{pnl2}.
The nearest substantial population centers to the site are the cities of Richland, Pasco,
and Kennewick, which in 2002 had a combined population of 124,000 \cite{census}.
The three cities are
20, 33, and 35 kilometers from the LIGO site. The LIGO Hanford laboratory is 
also located 15~km from sites 200E and 200W, facilities run by the Department of Energy
which are currently carrying out extensive environmental cleanup operations
which involve use of heavy earth moving machinery.
The Hanford site climatology is described in detail in \cite{pnl1}.
The average annual rainfall 18~cm. During the summer months, the site is
subject to strong day-night air temperature fluctuations. In July and August
the maximum daytime and minimum nighttime 
temperatures are typically $\mathrm{32~^\circ C}$ and $\mathrm{16~^\circ C}$.  
Strong wind gusts,
which can cause increased motion of the concrete slabs supporting the 
LIGO instruments, are observed. Peak wind gusts speeds in a one month
period are typically between $\mathrm{29~m/s}$ and $\mathrm{36~m/s}$. 
Gusts above $\mathrm{18~m/s}$ are more frequent in the months January
to April.

The Livingston site is located in managed pine forest and more general
forested land in southeastern Louisiana,  54~km from the nearest
point on Lake Ponchatrain, and 130~km from the nearest point on the Gulf of Mexico. 
The vertex latitude and longitude of the Livingston laboratory are
$30^{\circ}33'46.4''$ N, $90^{\circ}46'27.3''$ W, and the x arm
orientation is $252.3^{\circ}$ (WSW). The geology local to the site consists of
silt, sand, and mud alluvium, deposited by the Mississippi river. The bulk
of this material was deposited before 2 million years ago when the Gulf coast moved
south to approximately its present position along with the bulk of new Mississippi
silt deposits. The water table is less than a meter below the ground, so the interferometer 
buildings rest on earth banking, consisting of the local silt and sand, 
raised several meters above the natural ground level. The Livingston 
interferometer corner station is 7~km from the center of the town of Livingston, which
in 2000 had a population of 1,342 \cite{census}.
Climatological data is available from \cite{lsuclimatology}.
The average annual rainfall in Livingston Parish is 152~cm.
Violent rain and windstorms occur frequently, especially during August and September.
Winter minimum and summer maximum
temperatures are around $\mathrm{0~^\circ C}$ and $\mathrm{33~^\circ C}$
respectively. A particularly important environmental factor for seismic noise is
the use of local land for timber harvesting. The heavy machinery
used for logging adjacent to the site is a significant source of seismic noise
for the interferometer. A second important environmental factor affecting 
seismic noise at the instruments is the passage of cargo trains along the 
railroad track 7km to the south of the interferometer corner station. Large
trains that run past the site twice every evening currently disrupt the operation 
of the interferometer for around 40 minutes out of every day. For this paper
we do not attempt to separate data from the Livingston site into that taken during
the passage of trains and that without. We present instead a broad statistical 
overview of the seismic environment over a long time period.

\begin{figure}
\begin{center}
\input{blrmsdataflow_testmod.pstex_t}
\end{center}
\caption{\label{fig:blrmsdataflow}
Data flow from ground to blrms data file.
The filters used to band limit the decimated data
are autoregressive, meaning that they use
previous values of output data from the filter as well as 
input data to determine each new output data value.
These filters have the useful property that the 
reciprocal of the lower pass
band edge frequency of the filter can be longer than the time
duration of the data segment used to calculate the
filtered data output. The sampling rate in the decimated data
is chosen to be over twice the upper pass band edge
frequency of the filter. For example, the 0.1--0.3~Hz filter
uses input data decimated to a sampling rate of 1~Hz, and
8 samples each of input and previous output
data are used to determine each new filtered data point.
For this example, only 8 seconds worth of data is 
needed to implement the lower band edge at a frequency
of 1/10~Hz.
}
\end{figure}
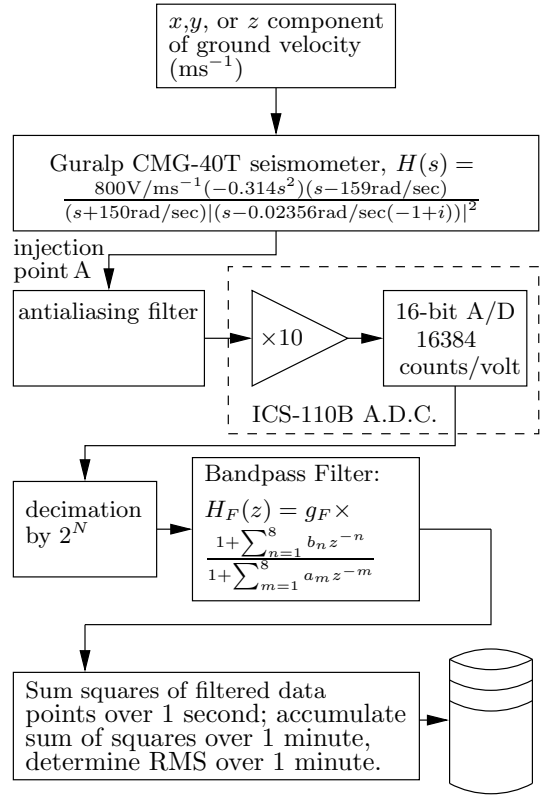

\begin{figure}
\begin{center}
\includegraphics[width=0.60\textwidth]{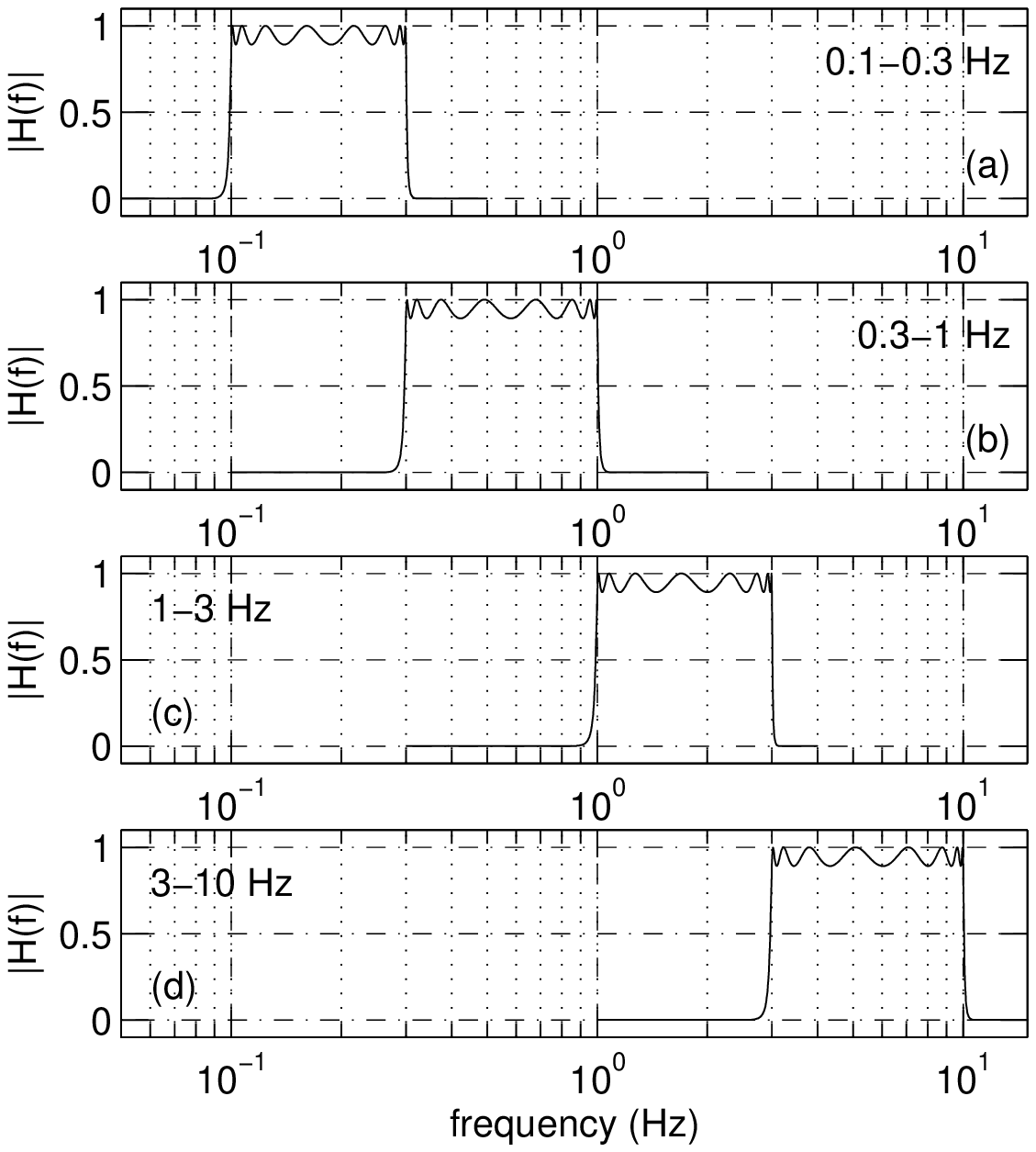}
\end{center}
\caption{\label{fig:filtertransfers}
Magnitudes of the transfer functions of the four time domain
bandpass filters applied to the seismometer data. The vertical axis
is the ratio of the amplitudes of the filter output to the filter input for a
test sinusoidal signal applied to the filter input, in the frequency range
indicated on the horizontal axis. Each filter is a 16th order elliptic type
having 1dB of passband ripple and 80dB of stopband attenuation.
}
\end{figure}

\section{Data Acquisition}
\label{sec:daq}

Data for this analysis were acquired using seismometers and acquisition 
hardware installed as part of the LIGO data acquisition system.
Figure \ref{fig:schematic} shows the approximate
locations of the 8 seismometers at the
two LIGO sites used in this experiment. The seismometers are Guralp CMG-40T
three-axis types \cite{guralp},
whose outputs are proportional to the three cartesian
components of slab velocity for disturbances in the frequency range
0.1--10~Hz. Figure \ref{fig:blrmsdataflow} is a schematic of the
data acquisition and the processing applied to the data. 
The voltage output of each seismometer
channel is passed through an antialiasing lowpass filter with a knee
frequency of 900~Hz, and is then digitized by an ICS-110B 16 bit A.D.C.
\cite{ics}
at a sampling rate of 2048Hz. The A.D.C. is housed in a V.M.E. crate
controlled by an 800MHz pentium-based CPU running VxWorks \cite{windriver},\
a real time operating system. A time stamp is added to the data
using a brandywine GPS receiver \cite{brandywine} installed in 
the V.M.E. crate.

The data acquired by the real time system are sent via fiber to a shared
memory partition on the backplane of a sun microsystems server which provides
an online data analysis environment \cite{daqfinaldesign} under UNIX.
Software running in this environment performs the analysis in
real time. The data are decimated to successively lower sampling rates, 
and passed through a bank of time-domain, 
bandpass filters. The filter frequency bands 
are 0.1--0.3~Hz, 0.3--1~Hz, 1--3~Hz, and 3--10~Hz. Figure
\ref{fig:filtertransfers} shows the magitudes of the filter transfer functions
for the four bands. 
The sum of the squares of the filtered data bins is measured once a second,
and 60 of these measurements taken in successive seconds are used to compute
the r.m.s.\ slab velocity at 1 minute intervals. The 
r.m.s.\ data are written to disk.

\section{Results}
\label{sec:results}
\subsection{The 0.1--0.3~Hz Frequency Band}

With the exception of short time periods during earthquakes, 
seismic noise in this band is dominated by `secondary microseisms'
\cite{cessaro}. They consist of travelling surface waves in the
earth's crust. The sources of these waves are complex, but they are
thought to originate in the pressure from counter-propagating ocean waves
against the ocean bottom near the coasts of the land masses, as
well as from more distant deep-ocean storms. Their frequency
content peaks in the range 0.05 - 0.5~Hz, but the position and shape
of the peak is highly non-stationary within this range. Large 
amplitude secondary microseisms are often observed during costal storms.

\begin{figure}
\begin{center}
\includegraphics[width=0.85\textwidth]{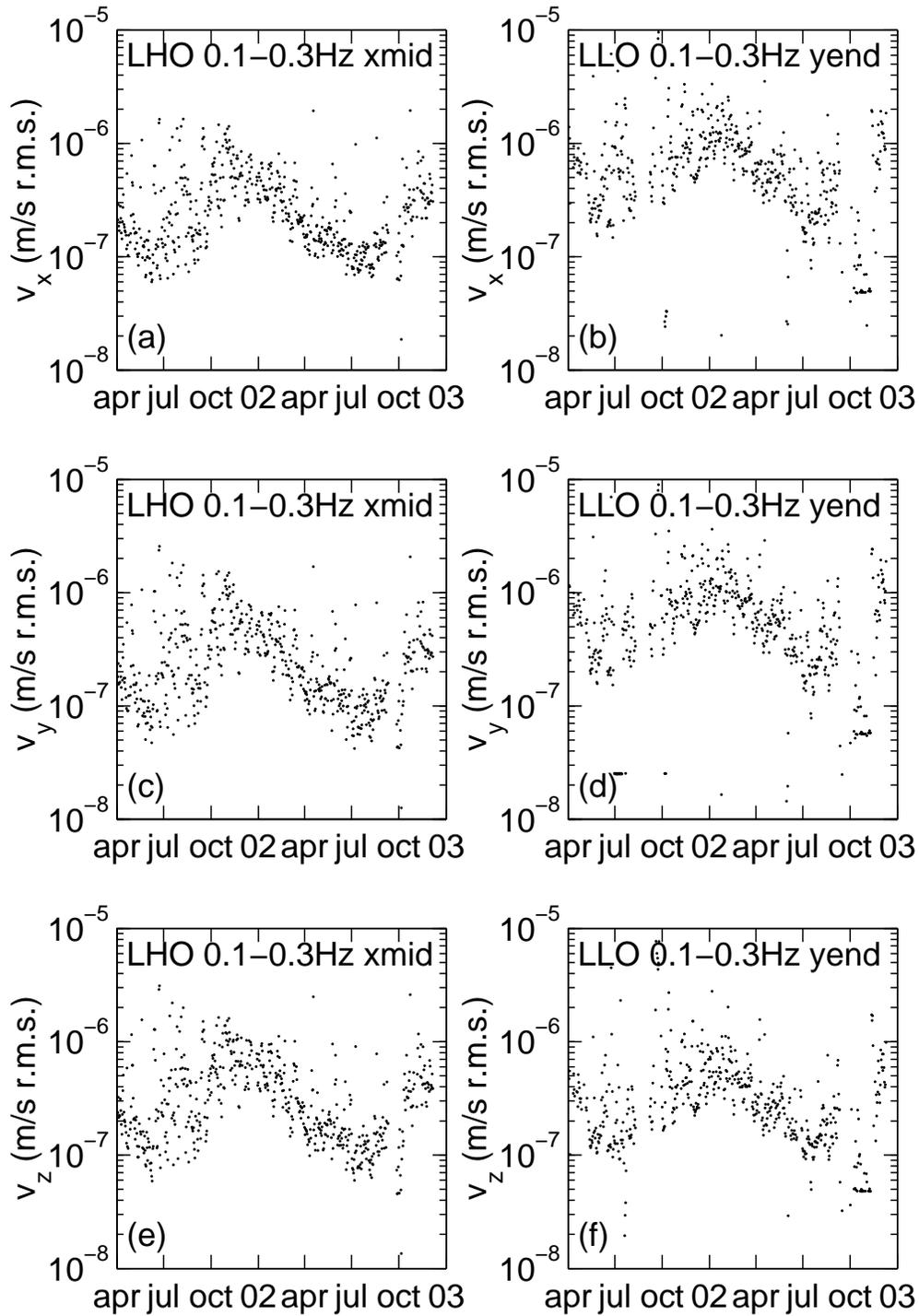}
\end{center}
\caption{\label{fig:p1p3perdayts}
Time series of the three cartesian components of ground velocity in 
$\mathrm{ms^{-1}}$ at the $x$ midstation at Hanford (left column)
and the $y$ endstation at Livingston (right column). The $x$ and $y$
components are parallel to the ground, the $z$ component is vertical.
Each point represents the r.m.s.\ in the frequency band 0.1--0.3~Hz
over one day. The independent variable ticks represent midnight on the
first of every third month, with the beginnings of years labelled by the
year number. These ticks are not evenly spaced because successive
three month intervals do not all contain the same number of days.
}
\end{figure}

\begin{figure}
\begin{center}
\includegraphics[width=0.95\textwidth]{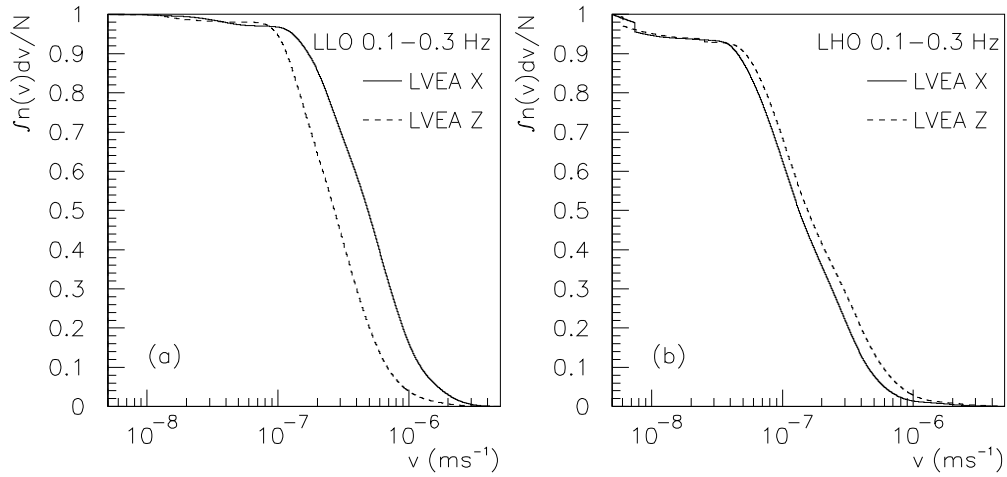}
\end{center}
\caption{\label{fig:cumulative_p1_p3_paper}
A cumulative histogram of r.m.s. velocity in the 0.1--0.3~Hz band.
The ordinate of the curve is the fraction of minute-duration data
segments having r.m.s. velocity greater than the abscissa,
The plot on the left gives results for a horizontal ($x$) and the
vertical ($z$) components measured by the Livingston corner station
seismometers. The right hand plot gives results for a horizontal ($x$) 
and the vertical ($z$) componenets measured by the Hanford corner
station seismometer. The number of points
in the histograms is 704150 (710212) for the Livingston horizontal (vertical) axis
histograms and 674286 (685362) for the Hanford horizontal (vertical) axis
histograms.
}
\end{figure}

\begin{figure}
\begin{center}
\includegraphics[width=0.95\textwidth]{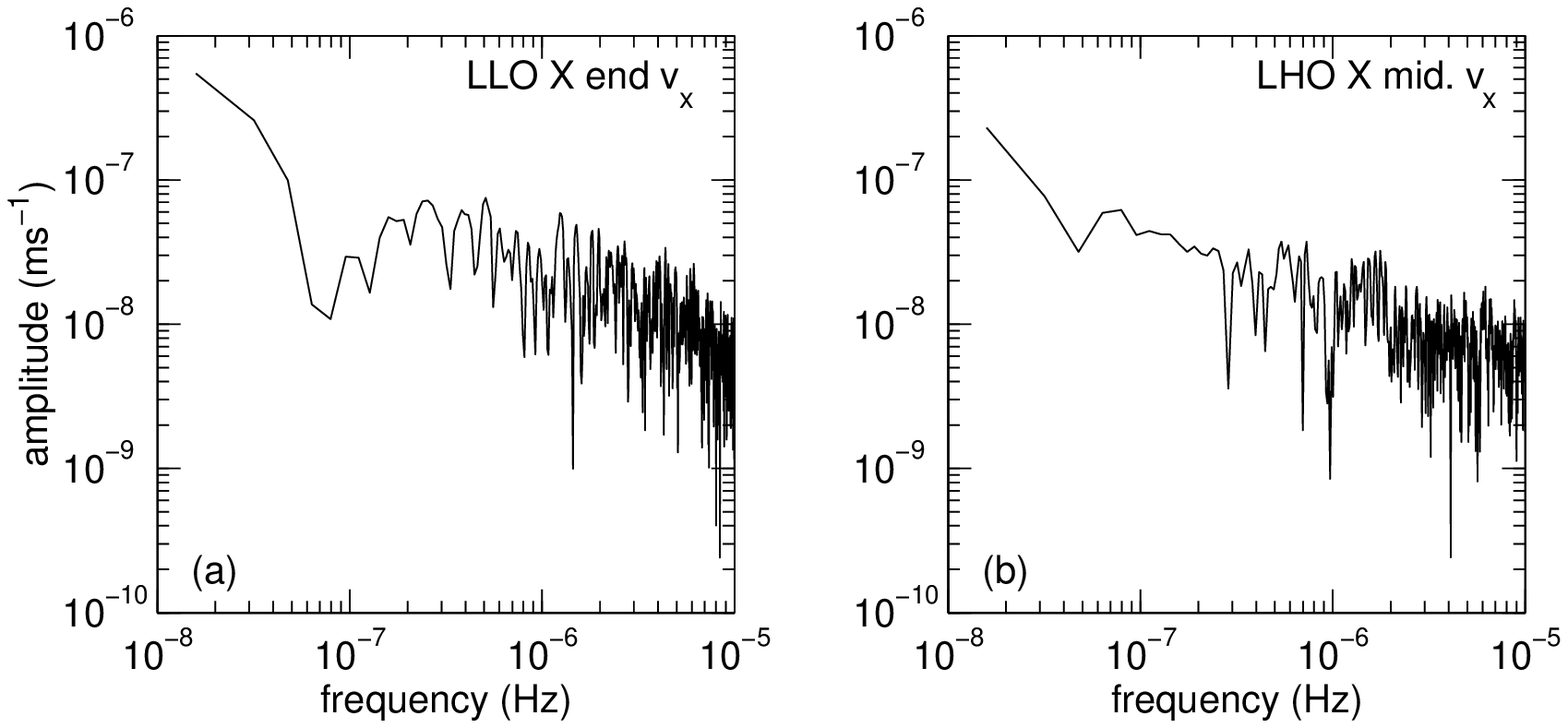}
\end{center}
\caption{\label{fig:p1p3psd}
Amplitude spectra of the time series output of
the 0.1--0.3~Hz filtered r.m.s.\ measurements from the $x$
axes horizontal output of the Livingston X endstation and 
Hanford X midstation seismometers.
}
\end{figure} 

Figure \ref{fig:p1p3perdayts} shows r.m.s.\ in the 0.1 - 0.3~Hz band over
613 days. All the data exhibit a marked annual modulation, with
the peaks of the modulation in the winter months and the troughs in the summer.
Figure \ref{fig:cumulative_p1_p3_paper} shows cumulative histograms of
all the velocity measurements taken over the 613 day period. Since the
two horizontal velocity components are similar, and we do not see significant
differences between the results for different seismometers at the same
LIGO site in this band, results are given for a single horizontal component
and vertical velocity for each site. Table \ref{tab:percentiles_p1_p3}
shows summary data for the 0.1--0.3~Hz band from each axis of 3 seismometers
from each site. The percentile values represent 
the velocities below which the given percentage of readings of the r.m.s.\ 
came over the 613 days of measurements.

Seismometers located in different buildings, but at the same site, gave
similar results in this frequency band. This is as expected, since the
microseismic wave amplitudes should not differ significantly over a 
distance of order 4km. The results in table \ref{tab:percentiles_p1_p3}, and
Figure \ref{fig:cumulative_p1_p3_paper} indicate that the Hanford site is
noisier than the Livingston site in this band by a factor of 2-3 in the horizontal
components. In the vertical components, the ratio is different at the 50th 
percentile level than at the 90th or 95th percentile level. The 50th percentile
noise levels are a factor of 1-2 higher at Livingston than at Hanford. At the 90th
and 95th percentile levels, a long tail in the cumulative statistics from Hanford
implies that 5 to 10 percent of the time the Hanford site is at least 80\% as noisy
as the Livingston site. 

It is interesting to note that the vertical components of seismic noise at Livingston are
less than the horizontal components by almost a factor of 2, at all percentile levels,
where at Hanford the vertical components equal or slightly exceed the horizontal
components.
Simplified theoretical models \cite{cessaro} treat the Earth's crust as an isotropic
half space of uniform composition, and contain two classes of traveling surface waves.
Rayleigh waves are elliptically polarized in the plane whose normal
is perpendicular to the wave vector and parallel to the Earth's surface, and
therefore produce both horizontal and vertical slab motion.
Love waves are linearly polarized parallel to the plane of the Earth's surface
and perpendicular to the direction of propagation of the wave. Therefore
Love waves produce only horizontal motion.
Assuming that the site slab motion can be decomposed into two components,
due to Rayleigh waves and Love waves, 
we conclude that differences in the sources and the geology through which the waves
have travelled to reach the sites cause the ellipticity of the
Rayleigh waves, the relative intensities of Rayleigh to Love waves, or
a combination of these factors, to differ between the sites. The data do not
allow us to make a direct comparison between the relative amplitudes of 
the Rayleigh and Love wave components at the two sites. 

Figure \ref{fig:p1p3psd} shows the amplitude spectral densities of the
time series of measurements of r.m.s.\ slab velocity in the $x$ axes
of the Livingston X end station and Hanford X mid station seismometers.
The annual modulation apparent from the time series of figure 
\ref{fig:p1p3perdayts} corresponds to a frequency of
$\mathrm{3.17\times 10^{-8}}$~Hz. The resolution bandwidth of the
power spectra are too low to resolve this peak from the zero frequency
component, so accurate estimates of the amplitude of annual modulation
of the microseism cannot be made without more data.
There is no evidence for other, shorter
periodicities. These amplitude spectra are typical of those from the full set
of seismometers and seismometer axes. The amplitude spectral density of 
slab motion fluctuations drop as the inverse square root of frequency 
starting at $\mathrm{4\times 10^{-7}}$~Hz, corresponding to a 28 day period.

\begin{table}
\caption{\label{tab:percentiles_p1_p3}
Table of 50th, 90th, and 95th percentile measured r.m.s.\ ground velocities
in three locations at each LIGO site, after applying a 
0.1--0.3~Hz bandpass filter. 
In the percentage of one minute measurements given at the head of
each column, the measured ground velocity in this band was less than
the tabulated value.} 
\begin{indented}
\item[]\begin{tabular}{
||p{18mm}|p{15mm}|p{7mm}|p{11mm}|p{11mm}|p{11mm}||} 
\br
\hline
\multicolumn{6}{||l||}{\bf 0.1--0.3~Hz band} \\ \hline
site & building & axis &
\multicolumn{3}{l||}
{velocity percentile ($\mu\rm m\,s^{-1}$)}
\\ \cline{4-6}
&&& 50\% & 90\% & 95\% \\ \hline
Livingston & Corner & $x$ & 0.49 & 1.2 & 1.7 \\
& & $y$ & 0.53 & 1.3 & 1.7 \\
& & $z$ & 0.27 & 0.69 & 0.91 \\  \cline{2-6}
& X end & $x$ & 0.50 & 1.2 & 1.7 \\
& & $y$ & 0.51 & 1.3 & 1.7 \\
& & $z$ & 0.25 & 0.67 & 0.89 \\  \cline{2-6}
& Y end & $x$ & 0.53 & 1.3 & 1.7 \\
& & $y$ & 0.50 & 1.3 & 1.6 \\
& & $z$ & 0.25 & 0.66 & 0.89 \\  \hline
Hanford & Corner & $x$ & 0.13 & 0.45 & 0.60 \\
& & $y$ & 0.12 & 0.43 & 0.58 \\
& & $z$ & 0.16 & 0.58 & 0.78 \\ \cline{2-6}
& X mid. & $x$ & 0.15 & 0.47 & 0.66 \\
& & $y$ & 0.13 & 0.46 & 0.64 \\
& & $z$ & 0.17 & 0.64 & 0.90 \\ \cline{2-6}
& Y mid. & $x$ & 0.15 & 0.49 & 0.67 \\
& & $y$ & 0.13 & 0.45 & 0.63 \\
& & $z$ & 0.13 & 0.45 & 0.63 \\ \hline
\br
\end{tabular}
\end{indented}
\end{table}

\subsection{The 0.3--1~Hz Frequency Band}

\begin{figure}
\begin{center}
\includegraphics[width=0.90\textwidth]{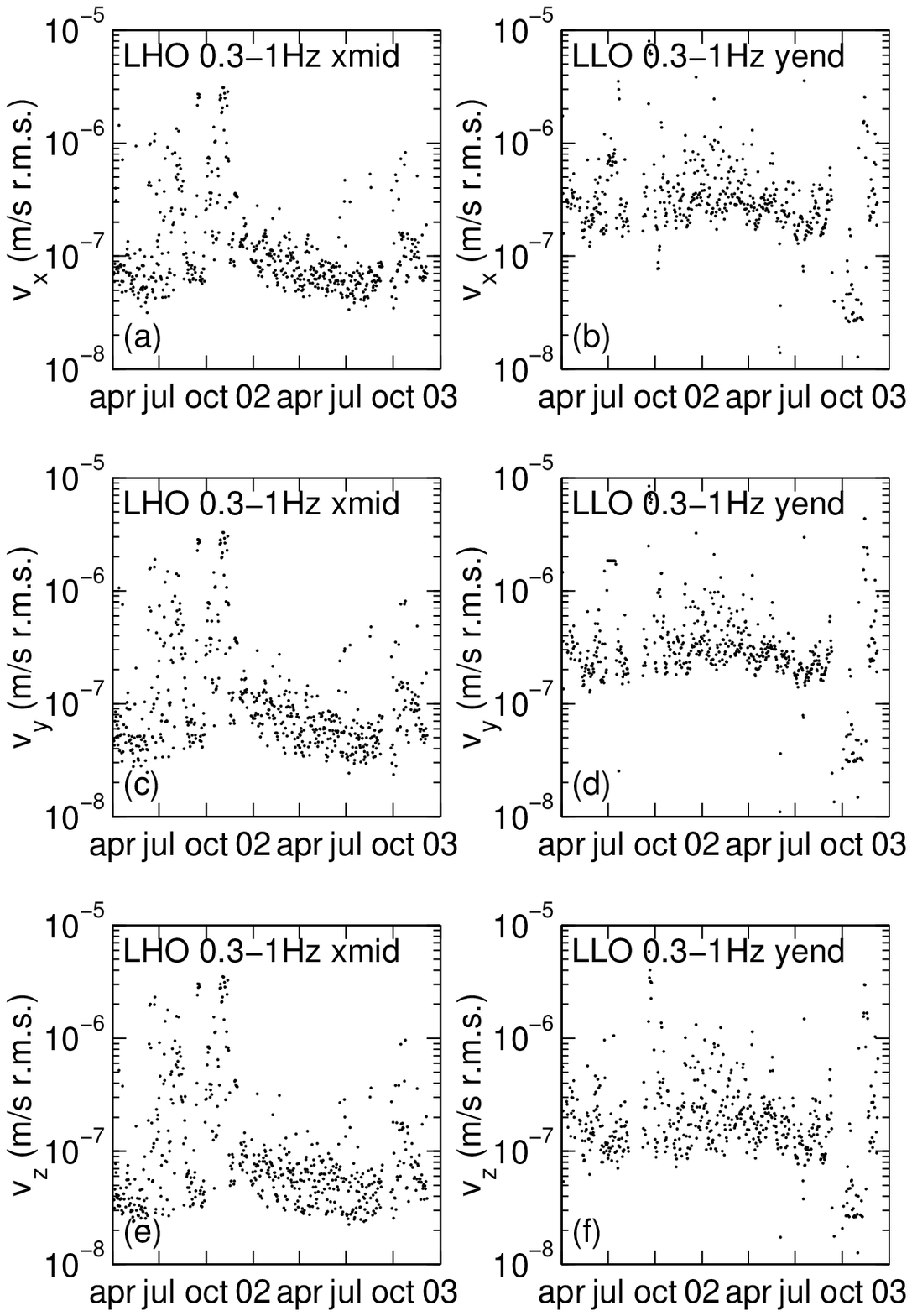}
\end{center}
\caption{\label{fig:p31onedayts}
Time series of the three cartesian components of ground velocity 
at the X midstation at Hanford (left column)
and the Y endstation at Livingston (right column) in the 
0.3--1~Hz frequency band. Each point is the r.m.s.\ over 1 day.
The labeling of the dependent variable axis on the plots
are the same as described in the caption of figure \ref{fig:p1p3perdayts}.
}
\end{figure} 

\begin{figure}
\begin{center}
\includegraphics[width=0.95\textwidth]{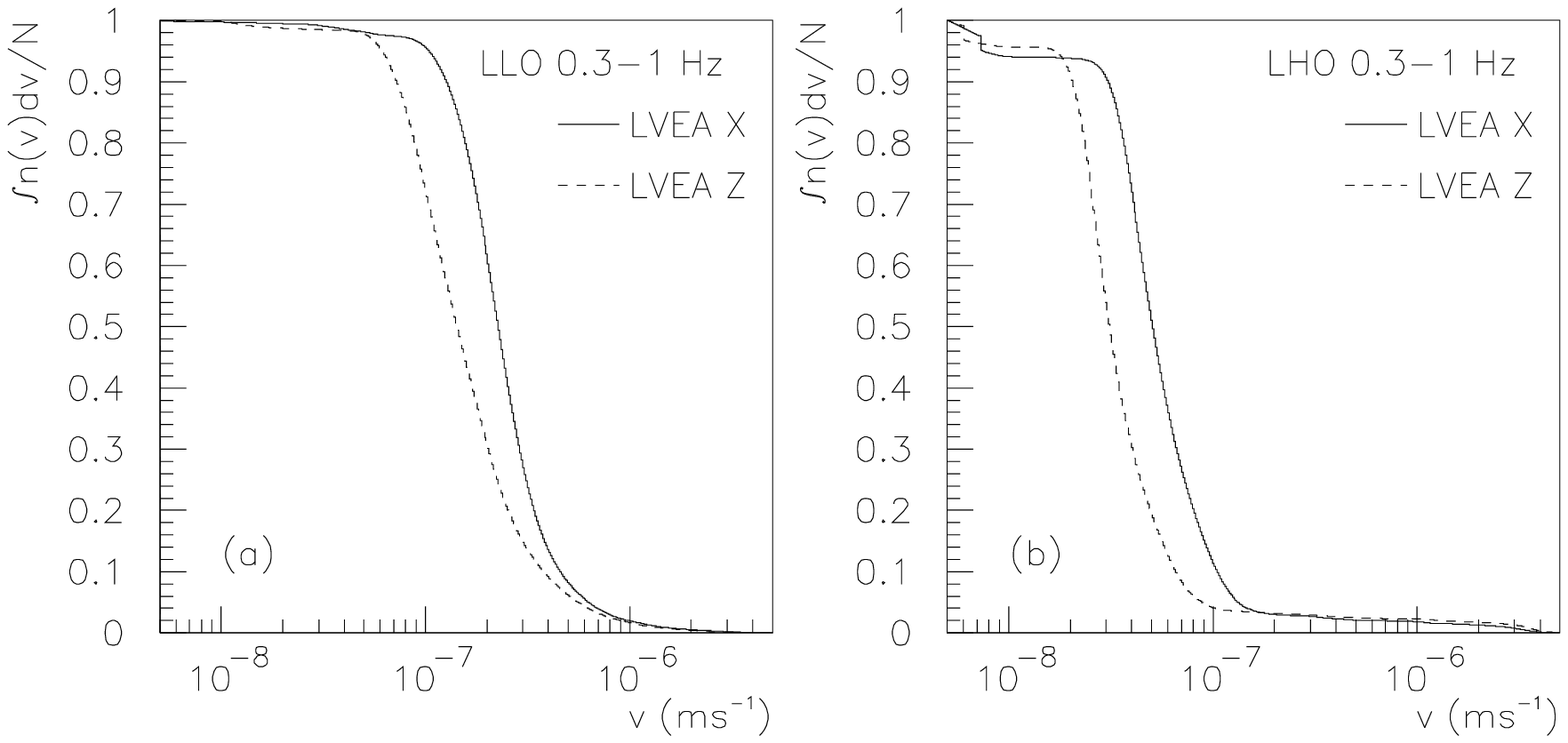}
\end{center}
\caption{\label{fig:cumplot_p3_1}
A cumulative histogram of the fraction of 1 minute duration measurements
of r.m.s.\ velocity in the 0.3--1~Hz band yielding a result greater than 
the abscissa. The results are for the same seismometers and directions
as described in the caption for figure \ref{fig:cumulative_p1_p3_paper}
The number of points in the histograms is 705302 (710434) for the 
Livingston horizontal (vertical) histograms and 678407 (678531) 
for the Hanford horizontal (vertical) histograms.
}
\end{figure}

Figure \ref{fig:p31onedayts} shows the three components of r.m.s.\
slab velocity in this band measured by the Hanford X midstation and
the Livingston Y endstation seismometers. 
Figure \ref{fig:cumplot_p3_1} gives cumulative histograms of the $x$ and
$z$ components of slab velocity in the Livingston and Hanford corner
stations, which were representative of the results from the other buildings.
Table \ref{tab:percentiles_p3_1} shows summary data for all seismometers
and axes, using the same percentile values as measured for the 
0.1--0.3~Hz band described above. 

There is some evidence for
a component of the slab motion having an annual modulation, which is
expected given that this frequency band encompasses the upper part
of the frequency band for secondary microseisms. The percentile velocities
from the table are a factor of 2--3 larger at Livingston than at Hanford. The ratio
of vertical to horizontal noise amplitude is about 1.5 for both sites. The two
components of horizontal motion at each seismometer have about the same
amplitude for each seismometer. 

The time series plots from Hanford show some large positive excursions, 
a factor of 10--30 in r.m.s. parallel to all axes between May and
October 2001. We speculate that these excursions may have been
caused by off site earth moving equipment near the Hanford interferometer,
although we are unable to provide further evidence in support of this
possibility. The Livingston time series data shows a period of lower amplitude in
September 2002. These are bad data due to some intrusive repairs and 
upgrades in the data acquisition system.

\begin{table}
\caption{\label{tab:percentiles_p3_1}
Table of 50th, 90th, and 95th percentile measured r.m.s.\ ground velocities
in three locations at each LIGO site, after applying a 
0.3--1~Hz bandpass filter. 
In the percentage of one minute measurements given at the head of
each column, the measured ground velocity in this band was less than
the tabulated value.}
\begin{indented}
\item[]\begin{tabular}{
||p{18mm}|p{15mm}|p{7mm}|p{11mm}|p{11mm}|p{11mm}||} 
\br
\hline
\multicolumn{6}{||l||}{\bf 0.3--1~Hz band} \\ \hline
site & building & axis &
\multicolumn{3}{l||}
{velocity percentile ($\mu\rm m\,s^{-1}$)}
\\ \cline{4-6}
&&& 50\% & 90\% & 95\% \\ \hline
Livingston & Corner & $x$ & 0.23 & 0.45 & 0.63 \\
& & $y$ & 0.22 & 0.45 & 0.63 \\
& & $z$ & 0.14 & 0.38 & 0.56 \\  \cline{2-6}
& X end & $x$ & 0.22 & 0.45 & 0.65 \\
& & $y$ & 0.21 & 0.44 & 0.63 \\
& & $z$ & 0.13 & 0.36 & 0.57 \\  \cline{2-6}
& Y end & $x$ & 0.25 & 0.59 & 0.83 \\
& & $y$ & 0.25 & 0.70 & 1.4 \\
& & $z$ & 0.15 & 0.41 & 0.61 \\  \hline
Hanford & Corner & $x$ & 0.051 & 0.10 & 0.13 \\
& & $y$ & 0.041 & 0.093 & 0.12 \\
& & $z$ & 0.031 & 0.066 & 0.088 \\ \cline{2-6}
& X mid. & $x$ & 0.060 & 0.12 & 0.17 \\
& & $y$ & 0.046 & 0.11 & 0.14 \\
& & $z$ & 0.035 & 0.071 & 0.098 \\ \cline{2-6}
& Y mid. & $x$ & 0.057 & 0.19 & 0.29 \\
& & $y$ & 0.044 & 0.15 & 0.22 \\
& & $z$ & 0.032 & 0.065 & 0.081 \\ \hline
\br
\end{tabular}
\end{indented}
\end{table}

\begin{figure}
\begin{center}
\includegraphics[width=0.65\textwidth]{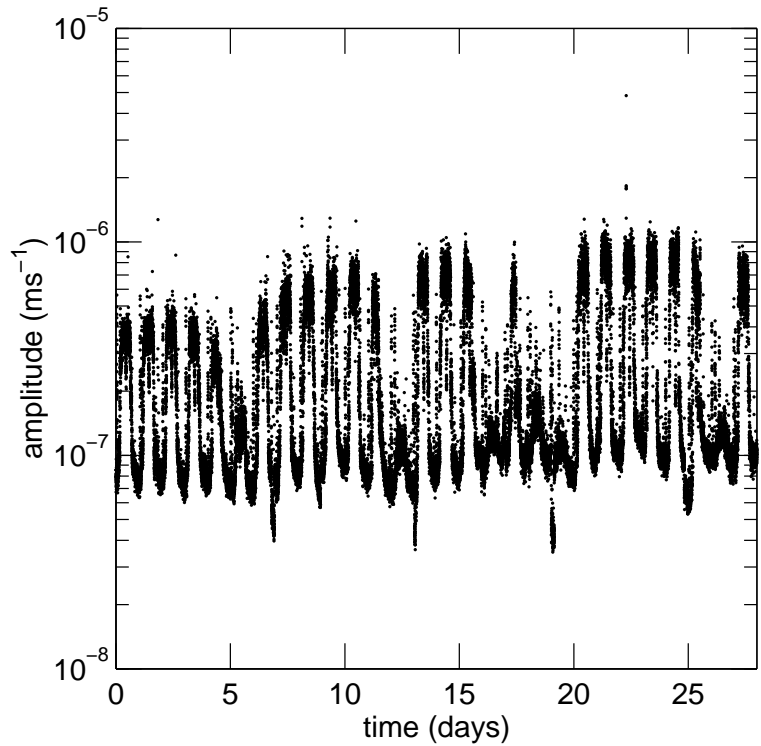}
\end{center}
\caption{\label{fig:thanksgiving}
Time series of r.m.s.\ noise in the 1--3~Hz band taken from 
a horizontal ($x$) axis of the Livingston X endstation seismometer. The
duration of the plot is 28 days. The spikes correspond to daytime,
the smaller spikes every 7 days are Sundays. The start time of the
plot is November 6th 2001, UTC 07:39:56. The large spikes are in 
groups of 6 for the weekdays plus Saturday, with the exception of the
3rd group for which the 4th spike is unusually small. This is
Thanksgiving day.
}
\end{figure}

\begin{figure}
\begin{center}
\includegraphics[width=0.90\textwidth]{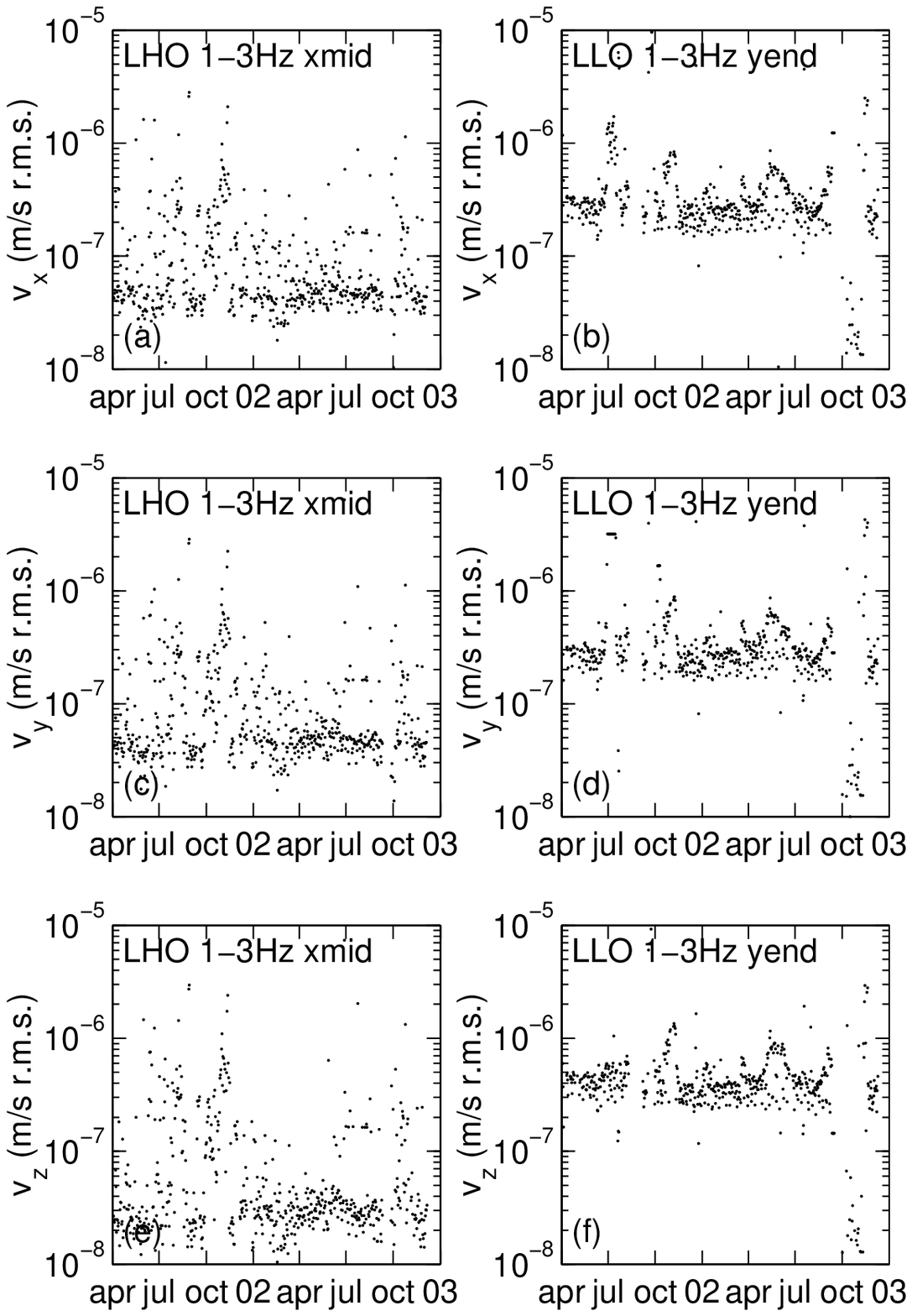}
\end{center}
\caption{\label{fig:13onedayts}
Time series of the three cartesian components of ground velocity
at the X midstation at Hanford (left column)
and the Y endstation at Livingston (right column) in the 
1--3~Hz frequency band. Each point is the r.m.s.\ over 1 day.
The labeling of the dependent variable axes
of the plots is the same as described in the
caption of figure \ref{fig:p1p3perdayts}.
}
\end{figure} 

\begin{figure}
\begin{center}
\includegraphics[width=0.95\textwidth]{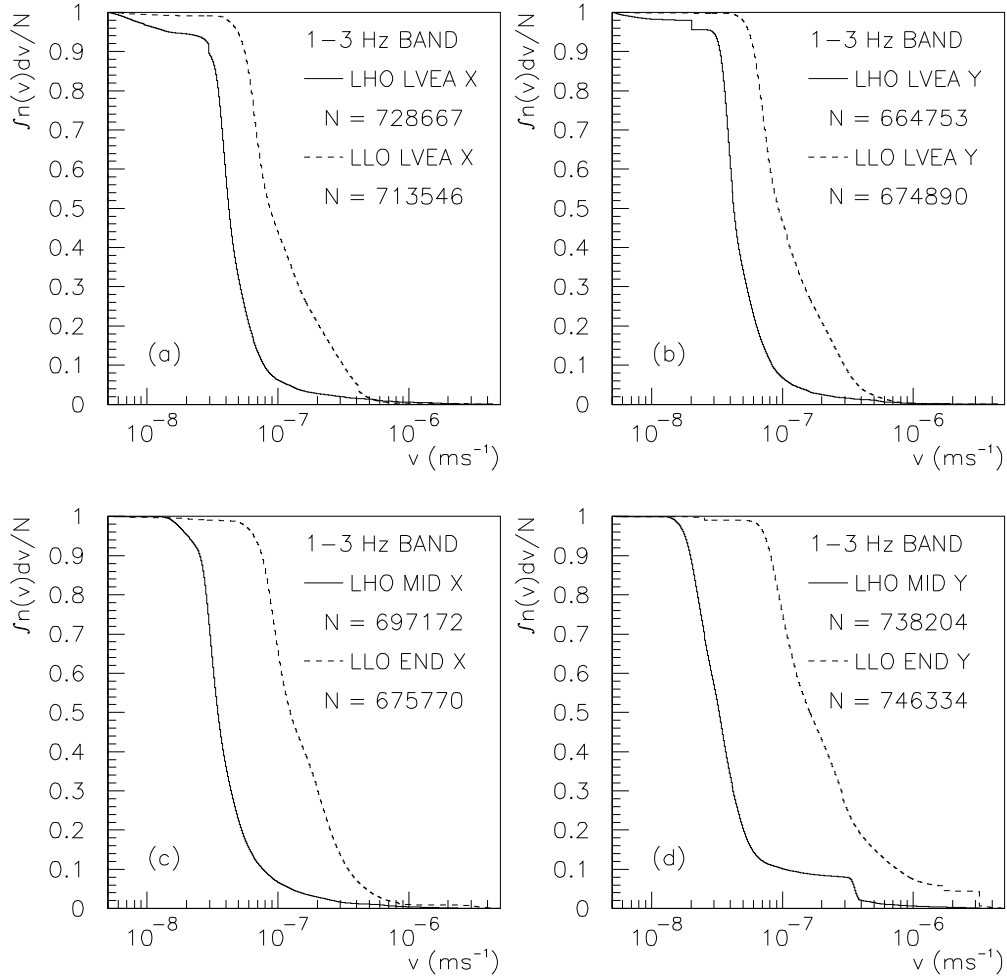}
\end{center}
\caption{\label{fig:inthist_1_3}
Cumulative histograms of the fraction of 1 minute duration measurements
of r.m.s.\ velocity in the 1--3~Hz band yielding a result greater than 
the abscissa. The four plots compare the slab velocity components at
parallel to the interferometer arms at the Livingston and Hanford corner
stations, and the same components at the Livingston end stations and the
Hanford mid stations. The plot legends give the number of events populating
the histograms.
}
\end{figure}

\begin{figure}
\begin{center}
\includegraphics[width=0.95\textwidth]{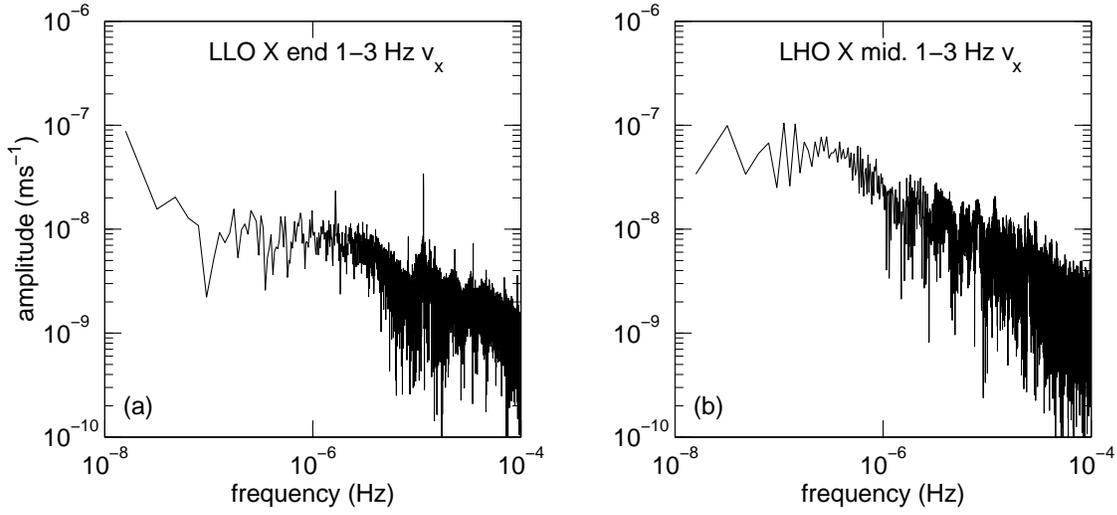}
\end{center}
\caption{\label{fig:13psd}
Amplitude spectra of the time series output of
the 1--3~Hz filtered r.m.s.\ measurements from a horizontal
($x$) component of the Livingston X endstation seismometer
and a horizontal ($x$) component of the Hanford X midstation
seismometers.}
\end{figure} 

\begin{table}
\caption{\label{tab:percentiles_1_3}
Table of 50th, 90th, and 95th percentile measured r.m.s.\ ground velocities
in three locations at each LIGO site, after applying a 
1--3~Hz bandpass filter.
In the percentage of one minute measurements given at the head of
each column, the measured ground velocity in this band was less than
the tabulated value.} 
\begin{indented}
\item[]\begin{tabular}{
||p{18mm}|p{15mm}|p{7mm}|p{11mm}|p{11mm}|p{11mm}||} 
\br
\hline
\multicolumn{6}{||l||}{\bf 1--3~Hz band} \\ \hline
site & building & axis &
\multicolumn{3}{l||}
{velocity percentile ($\mu\rm m\,s^{-1}$)}
\\ \cline{4-6}
&&& 50\% & 90\% & 95\% \\ \hline
Livingston & Corner & $x$ & 0.087 & 0.32 & 0.40 \\
& & $y$ & 0.093 & 0.30 & 0.38 \\
& & $z$ & 0.12 & 0.56 & 0.75 \\  \cline{2-6}
& X end & $x$ & 0.12 & 0.34 & 0.47 \\
& & $y$ & 0.21 & 0.40 & 0.53 \\
& & $z$ & 0.13 & 0.39 & 0.55 \\  \cline{2-6}
& Y end & $x$ & 0.16 & 0.65 & 1.1 \\
& & $y$ & 0.16 & 0.75 & 1.7 \\
& & $z$ & 0.20 & 0.72 & 1.1 \\  \hline
Hanford & Corner & $x$ & 0.042 & 0.079 & 0.12 \\
& & $y$ & 0.042 & 0.083 & 0.12 \\
& & $z$ & 0.020 & 0.059 & 0.097 \\ \cline{2-6}
& X mid. & $x$ & 0.035 & 0.077 & 0.13 \\
& & $y$ & 0.033 & 0.077 & 0.12 \\
& & $z$ & 0.019 & 0.045 & 0.077 \\ \cline{2-6}
& Y mid. & $x$ & 0.037 & 0.15 & 0.31 \\
& & $y$ & 0.032 & 0.10 & 0.36 \\
& & $z$ & 0.033 & 0.10 & 0.36 \\ \hline
\br
\end{tabular}
\end{indented}
\end{table}

\subsection{The 1--3~Hz Frequency Band}

Slab motion in the 1--3~Hz band contains no appreciable contribution from
secondary microseisms. The dominant contribution to slab motion in this
band derives from human activies involving vehicles and machinery in 
the vicinity of the sites. Figure \ref{fig:thanksgiving} shows one minute
measurements of the $x$ component of slab velocity in the Livingston Y
endstation over 28 days commencing at 07:39:56 UTC on November 6th 2001.
The measured velocity is dominated by a strong day night modulation whose
amplitude decreases for one day per week. The higher noise level in the
daytime is due to human activity. The lower amplitude of the daytime peak
is on Sundays. The 16th day of this data set
was Thanksgiving, which has a lower level of daytime slab noise than on
the other weekdays.

This frequency band is particularly important to the interferometers, since
two of the resonances of the vibration isolation stack under each LIGO 
optic are at 1.2~Hz and 2.1~Hz. Excess ground
motion in this frequency range can ring up these resonances, adding to the
force noise on the mirrors, which increases the fraction of the dynamic
reserve of the control systems utilized in surpressing mirror motions. 
By this mechanism, 1--3~Hz ground motion often causes lock loss, 
particularly at Livingston. Also, even when lock is not lost, the resultant
increased r.m.s. deviation from the dark fringe adds shot noise and
increases the sensitivity to other technical noises. 

Figure \ref{fig:13onedayts} shows time series of the cartesian components
of slab velocity for one seismometer at each site. There is no evidence
for annual modulation. Because each data point on these
plots represents an r.m.s.\ reading over a whole day, the modulation during
the day due to human activity is invisible. On the noisiest days at 
Livingston, the slab velocity averages $\mathrm{3~\mu m/s}$. 
There are also obvious periods of increased human activity lasting
on the order of a month. It is now known that the Livingston excess noise
is caused primarily by logging activities in the vicinity of the site. There
is also a contribution from the town of Livingston. There is no evidence
here for a trend towards increasing r.m.s. slab velocity over
the 613 days of this measurement.

Figure \ref{fig:inthist_1_3} shows cumulative histograms of 1--3~Hz
seismic noise. The four plots give results from the two seismometer
axes parallel to each interferometer arm at each site. Table 
\ref{tab:percentiles_1_3} gives percentile velocities in all seismometers
and axes. The noisiest
station is the Y endstation at Livingston. Measuring parallel to the
beam pipe, the 90th percentile slab velocity is 0.75, a factor of 
5 greater than the noisiest data from Hanford, which is the $x$ axis
of the Y midstation velocity. The ratio of the 90th percentile velocities
of the noisiest horizontal Livingston seismometer axis to the quietest 
horizontal Hanford seismometer axis is about 10.

Figure \ref{fig:13psd} shows amplitude spectra from the $x$ axes of the 
Livingston X endstation and the Hanford X midstation. The strong peaks
at frequencies of $\mathrm{12 \mu Hz}$ and $\mathrm{1.6 \mu Hz}$ 
correspond to periods of 1 day and 1 week respectively. One day and one
week periodicities are not evident in the data fron the Hanford X 
midstation. This is the seismometer that is furthest from the roads
out of the three used in this study. Some day-night variation can 
be seen in time series from the Y midstation, but it is much weaker
than that seen at Livingston.

\begin{figure}
\begin{center}
\includegraphics[width=0.90\textwidth]{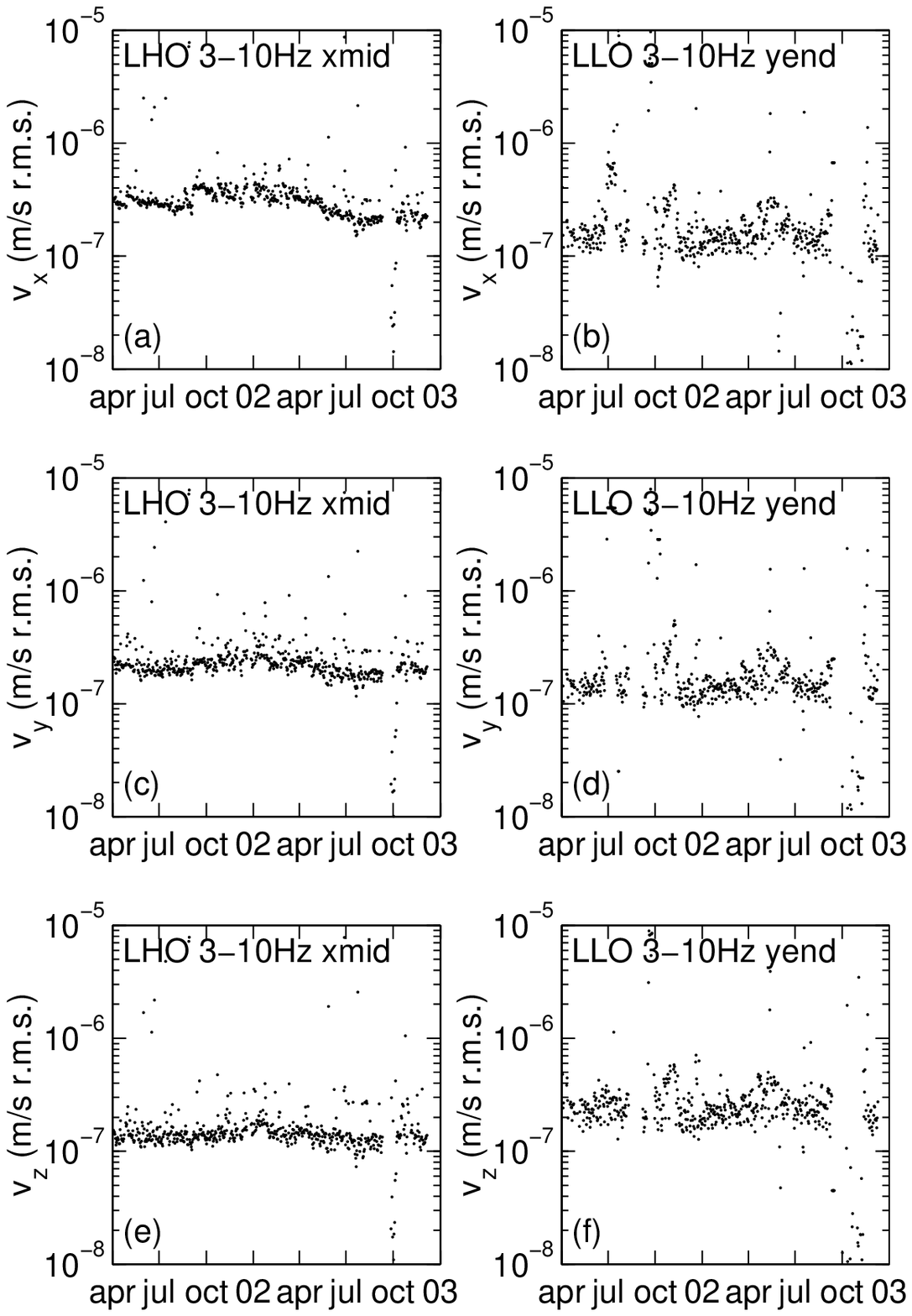}
\end{center}
\caption{\label{fig:310onedayts}
Time series of the three cartesian components of ground velocity 
at the X midstation at Hanford (left column)
and the Y endstation at Livingston (right column) in the 
3--10~Hz frequency band. Each point is the r.m.s.\ over 1 day.
The labeling of the dependent variable of the plots
is the same as described in the
caption of figure \ref{fig:p1p3perdayts}.
}
\end{figure} 

\begin{figure}
\begin{center}
\includegraphics[width=0.95\textwidth]{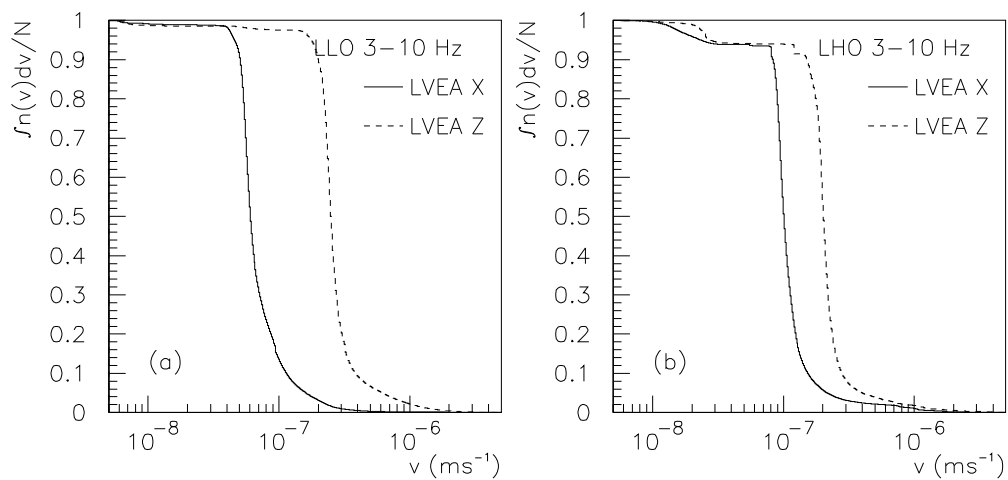}
\end{center}
\caption{\label{fig:cumplot_3_10}
A cumulative histogram of the fraction of 1 minute duration measurements
of r.m.s.\ velocity in the 3--10~Hz band yielding a result greater than 
the abscissa. The plot on the left gives results for a horizontal ($x$) and 
the vertical ($z$) component measured in the Livingston corner station. 
The plot on the right gives results for a horizontal ($x$) and the vertical
($z$) component at the Hanford corner station. 
The number of points in the histograms is 726730 (727500) for the 
Livingston horizontal (vertical) axis
histograms and 699145 (703879) for the Hanford horizontal (vertical)
axis histograms.
}
\end{figure}

\begin{figure}
\begin{center}
\includegraphics[width=0.95\textwidth]{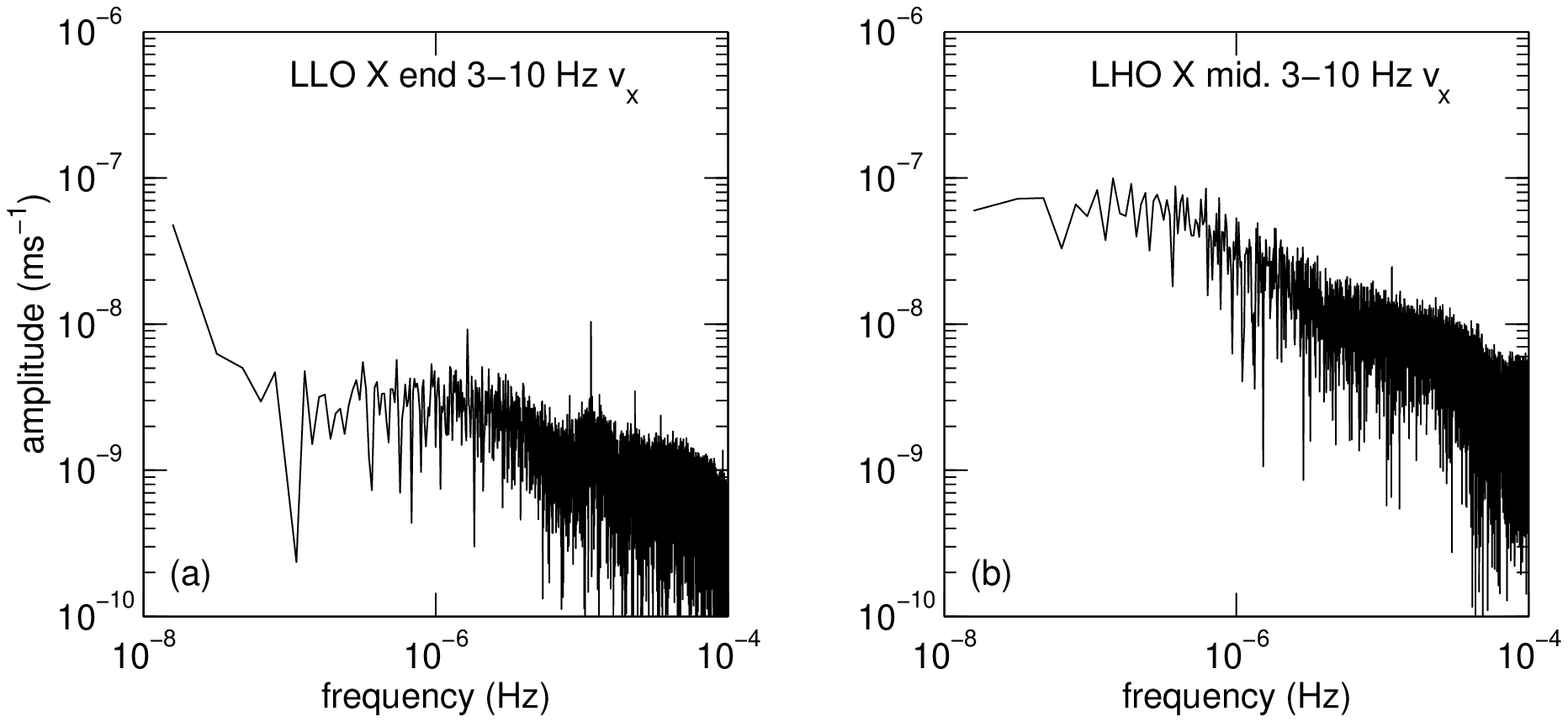}
\end{center}
\caption{\label{fig:310psd}
Amplitude spectra of the time series output of
the 3--10~Hz filtered r.m.s.\ measurements for a horizontal ($x$)
axis of the Livingston X endstation and the Hanford X midstation
seismometers.}
\end{figure}

\subsection{The 3--10~Hz Frequency Band}

Like the 1--3~Hz band, the 3--10~Hz encompasses a resonance, at
6.3~Hz, of the vibration isolation stack. 
This frequency band shows the day night modulation associated with human
activity, but at a lower level than was seen in the 1--3~Hz band. Figure
\ref{fig:310onedayts} shows r.m.s.\ measured over one day of 3--10~Hz
slab velocity in the Hanford X midstation and the Livingston Y endstation.
Table \ref{tab:percentiles_3_10} shows the 50th, 90th, and 95th percentile
slab velocities in all seismometers and axes. Figure
\ref{fig:cumplot_3_10} shows the cumulative histograms of slab
velocities at the Livingston and Hanford corner station $x$ and $z$
axes.

In the 3--10~Hz band, slab velocities at Hanford are 5--50~\% higher than
at Livingston. There is much larger variation between the slab velocities
measured in different buildings at the same site than for the other bands.
We also see greater difference between measured velocity percentiles along
different axes of the same seismometer, suggesting that specific close
sources of vibration might be the cause of ground motion at these frequencies.
At Livingston, in contrast to the other frequency bands, the vertical
component of slab velocity is at least as large as the horizontal components.
The most extreme case, the Livingston corner station,  
is illustrated in figure \ref{fig:cumplot_3_10}, where the $z$ component
of slab motion is three times higher than the $x$ and $y$ components. An 
anomolously large vertical slab motion in the corner station is also seen
at Hanford. This pattern suggests that some local source of vibration
is exciting the slab in the corner stations at both sites. The corner 
buildings contain more mechanical equipment than the arm buildings.
We do not think it likely that this effect is due to movement and activity
of the LIGO scientists in the corner stations. In our experience,
human activities would also lead to an excess in the 
1--3~Hz band, which is not seen. 

\begin{table}
\caption{\label{tab:percentiles_3_10}
Table of 50th, 90th, and 95th percentile measured r.m.s.\ ground velocities
in three locations at each LIGO site, after applying a 
3--10~Hz bandpass filter.
In the percentage of one minute measurements given at the head of
each column, the measured ground velocity in this band was less than
the tabulated value.} 
\begin{indented}
\item[]\begin{tabular}{
||p{18mm}|p{15mm}|p{7mm}|p{11mm}|p{11mm}|p{11mm}||}
\br
\hline
\multicolumn{6}{||l||}{\bf 3--10~Hz band} \\ \hline
site & building & axis &
\multicolumn{3}{l||}
{velocity percentile ($\mu\rm m\,s^{-1}$)}
\\ \cline{4-6}
&&& 50\% & 90\% & 95\% \\ \hline
Livingston & Corner & $x$ & 0.060 & 0.11 & 0.16 \\
& & $y$ & 0.081 & 0.13 & 0.20 \\
& & $z$ & 0.25 & 0.38 & 0.61 \\  \cline{2-6}
& X end & $x$ & 0.15 & 0.24 & 0.33 \\
& & $y$ & 0.22 & 0.30 & 0.39 \\
& & $z$ & 0.25 & 0.38 & 0.46 \\  \cline{2-6}
& Y end & $x$ & 0.094 & 0.34 & 0.52 \\
& & $y$ & 0.099 & 0.28 & 0.45 \\
& & $z$ & 0.15 & 0.39 & 0.56 \\  \hline
Hanford & Corner & $x$ & 0.10 & 0.15 & 0.21 \\
& & $y$ & 0.11 & 0.17 & 0.23 \\
& & $z$ & 0.20 & 0.28 & 0.40 \\ \cline{2-6}
& X mid. & $x$ & 0.29 & 0.41 & 0.44 \\
& & $y$ & 0.20 & 0.28 & 0.32 \\
& & $z$ & 0.12 & 0.17 & 0.21 \\ \cline{2-6}
& Y mid. & $x$ & 0.15 & 0.26 & 0.57 \\
& & $y$ & 0.19 & 0.28 & 0.66 \\
& & $z$ & 0.14 & 0.19 & 0.21 \\ \hline
\br
\end{tabular}
\end{indented}
\end{table}

\section{Conclusions and Future Work}
\label{sec:conclusion}

We have presented a detailed study of the ground noise properties of
the slabs supporting the LIGO vacuum system and optical components.
For the lower frequency bands investigated, between 0.1 and 1 Hz, the
noise is dominated by microseismic waves. Between 1 and 10 Hz, the
dominant noise sources are due to human activity in the vicinity of
the instruments.

The most critical band for operation of the interferometer is the 1--3~Hz
band, in which the r.m.s.\ slab velocity parallel to the interferometer 
arms is typically four times as high at Livingston as at Hanford. The
noise level increases in the daytime, and is due to human activity,
particularly logging machinery close to the Livingston site. We
see no long term trend of increasing slab velocity over the 613 day 
duration of these measurements at Livingston. Slab velocity in the
Livingston Y 
endstation is a factor of 2 to 2.5 larger than in the X endstation or
the corner station. Seismic filtering retrofits \cite{retrofits}
should take into account
the particularly noisy environment in this building.

In the 3--10~Hz band the effects of human activity are still seen, but
are typically at a lower level than in the 1--3~Hz band. Slab noise
parallel to the ground appears less isotropic than in the 1--3~Hz band,
and slab velocities are typically larger in the arm buildings than
the corner stations at both sites.

In both the 1--3~Hz and 3--10~Hz bands at Livingston, 
the vertical component of slab velocity is equal to or larger than the 
horizontal components. For these same bands at Hanford, the
vertical component is larger in the corner stations but smaller in the
arm buildings than the horizontal components. 

The two frequency bands below 1~Hz are dominated by the secondary 
microseism. We see evidence for annual modulation of the slab velocity
in this band. A longer data set is needed to determine the amplitude
and stationarity of this annual modulation. The ratio of vertical to
horizontal slab motion at Livingston is 0.5 to 0.7 in the 0.1--0.3~Hz
band, whereas at Hanford this ratio is 1--1.3. In the 0.3--1~Hz band,
the 90th percentile vertical component of slab velocity is typically 
smaller than the horizontal components at both sites.

Future enhancements of this analysis are planned. A
detailed study of the ratio of peak slab velocities to the r.m.s.\ levels
determined in this study would be useful in
designing upgrades to improve isolation of the LIGO optics from the
ground. A specialized study focusing on ground motion in the
1--3~Hz frequency band at Livingston during train passages
is planned, and would provide useful data for instrument upgrades
designed to improve interferometer isolation from ground disturbances.

\ack

This work was supported by National Science Foundation awards 9801158,
0107417, and 0071316, as well as Louisiana Board of Regents contract
LEQSF(2000-03)-RD-A-06. We wish to thank Gabriela Gonzalez, Peter Fritschel,
and Robert Schofield for useful discussions and comments,
and the the scientific and technical staff of LIGO for commissioning and
maintaining the interferometers and data acqusition system.

\end{document}

%% file: bothsites_horiz.pstex_t
\begin{picture}(0,0)%
\includegraphics{bothsites_horiz.pstex}%
\end{picture}%
\setlength{\unitlength}{3947sp}%
\begingroup\makeatletter\ifx\SetFigFont\undefined%
\gdef\SetFigFont#1#2#3#4#5{%
  \reset@font\fontsize{#1}{#2pt}%
  \fontfamily{#3}\fontseries{#4}\fontshape{#5}%
  \selectfont}%
\fi\endgroup%
\begin{picture}(3764,7520)(399,-6795)
\put(2866,-3782){\makebox(0,0)[lb]{\smash{{\SetFigFont{9}{10.8}{\familydefault}{\mddefault}{\updefault}{\color[rgb]{0,0,0}BEAMSPLITTER}%
}}}}
\put(1761,-1638){\makebox(0,0)[lb]{\smash{{\SetFigFont{9}{10.8}{\familydefault}{\mddefault}{\updefault}{\color[rgb]{0,0,0}B}%
}}}}
\put(1871,-2118){\makebox(0,0)[lb]{\smash{{\SetFigFont{9}{10.8}{\familydefault}{\mddefault}{\updefault}{\color[rgb]{0,0,0}M}%
}}}}
\put(1871,-2378){\makebox(0,0)[lb]{\smash{{\SetFigFont{9}{10.8}{\familydefault}{\mddefault}{\updefault}{\color[rgb]{0,0,0}M}%
}}}}
\put(1731,-3420){\makebox(0,0)[lb]{\smash{{\SetFigFont{9}{10.8}{\familydefault}{\mddefault}{\updefault}{\color[rgb]{0,0,0}Y END}%
}}}}
\put(1656,-4920){\makebox(0,0)[lb]{\smash{{\SetFigFont{9}{10.8}{\familydefault}{\mddefault}{\updefault}{\color[rgb]{0,0,0}CORNER}%
}}}}
\put(2536,-3387){\makebox(0,0)[lb]{\smash{{\SetFigFont{9}{10.8}{\familydefault}{\mddefault}{\updefault}{\color[rgb]{0,0,0}M}%
}}}}
\put(2566,-3872){\makebox(0,0)[lb]{\smash{{\SetFigFont{9}{10.8}{\familydefault}{\mddefault}{\updefault}{\color[rgb]{0,0,0}B}%
}}}}
\put(2866,-3392){\makebox(0,0)[lb]{\smash{{\SetFigFont{9}{10.8}{\familydefault}{\mddefault}{\updefault}{\color[rgb]{0,0,0}MIRROR}%
}}}}
\put(1811,-6050){\makebox(0,0)[lb]{\smash{{\SetFigFont{9}{10.8}{\familydefault}{\mddefault}{\updefault}{\color[rgb]{0,0,0}M}%
}}}}
\put(3756,-5520){\makebox(0,0)[lb]{\smash{{\SetFigFont{9}{10.8}{\familydefault}{\mddefault}{\updefault}{\color[rgb]{0,0,0}X END}%
}}}}
\put(1291,-5510){\makebox(0,0)[lb]{\smash{{\SetFigFont{9}{10.8}{\familydefault}{\mddefault}{\updefault}{\color[rgb]{0,0,0}M}%
}}}}
\put(1131,-3340){\makebox(0,0)[lb]{\smash{{\SetFigFont{9}{10.8}{\familydefault}{\mddefault}{\updefault}{\color[rgb]{0,0,0}M}%
}}}}
\put(3991,-6200){\makebox(0,0)[lb]{\smash{{\SetFigFont{9}{10.8}{\familydefault}{\mddefault}{\updefault}{\color[rgb]{0,0,0}M}%
}}}}
\put(1031,-6330){\makebox(0,0)[lb]{\smash{{\SetFigFont{9}{10.8}{\familydefault}{\mddefault}{\updefault}{\color[rgb]{0,0,0}B}%
}}}}
\put(2811,-6660){\makebox(0,0)[lb]{\smash{{\SetFigFont{12}{14.4}{\familydefault}{\mddefault}{\updefault}{\color[rgb]{0,0,0}LIVINGSTON}%
}}}}
\put(851,-2163){\makebox(0,0)[lb]{\smash{{\SetFigFont{9}{10.8}{\familydefault}{\mddefault}{\updefault}{\color[rgb]{0,0,0}M}%
}}}}
\put(1881,-1643){\makebox(0,0)[lb]{\smash{{\SetFigFont{9}{10.8}{\familydefault}{\mddefault}{\updefault}{\color[rgb]{0,0,0}M}%
}}}}
\put(1001,-2428){\makebox(0,0)[lb]{\smash{{\SetFigFont{9}{10.8}{\familydefault}{\mddefault}{\updefault}{\color[rgb]{0,0,0}B}%
}}}}
\put(861,-6058){\makebox(0,0)[lb]{\smash{{\SetFigFont{9}{10.8}{\familydefault}{\mddefault}{\updefault}{\color[rgb]{0,0,0}M}%
}}}}
\put(961,-1618){\makebox(0,0)[lb]{\smash{{\SetFigFont{9}{10.8}{\familydefault}{\mddefault}{\updefault}{\color[rgb]{0,0,0}M}%
}}}}
\put(3769,-1192){\rotatebox{234.0}{\makebox(0,0)[lb]{\smash{{\SetFigFont{9}{10.8}{\rmdefault}{\mddefault}{\updefault}{\color[rgb]{0,0,0}N}%
}}}}}
\put(2394,-1590){\makebox(0,0)[lb]{\smash{{\SetFigFont{9}{10.8}{\familydefault}{\mddefault}{\updefault}{\color[rgb]{0,0,0}Z}%
}}}}
\put(2634,-1088){\makebox(0,0)[lb]{\smash{{\SetFigFont{9}{10.8}{\familydefault}{\mddefault}{\updefault}{\color[rgb]{0,0,0}Y}%
}}}}
\put(2949,-1365){\makebox(0,0)[lb]{\smash{{\SetFigFont{9}{10.8}{\familydefault}{\mddefault}{\updefault}{\color[rgb]{0,0,0}X}%
}}}}
\put(2791,-2761){\makebox(0,0)[lb]{\smash{{\SetFigFont{12}{14.4}{\familydefault}{\mddefault}{\updefault}{\color[rgb]{0,0,0}HANFORD}%
}}}}
\put(2941,-646){\makebox(0,0)[lb]{\smash{{\SetFigFont{9}{10.8}{\familydefault}{\mddefault}{\updefault}{\color[rgb]{0,0,0}SEISMOMETER}%
}}}}
\put(2941,-256){\makebox(0,0)[lb]{\smash{{\SetFigFont{9}{10.8}{\familydefault}{\mddefault}{\updefault}{\color[rgb]{0,0,0}BUILDING}%
}}}}
\put(2941,119){\makebox(0,0)[lb]{\smash{{\SetFigFont{9}{10.8}{\familydefault}{\mddefault}{\updefault}{\color[rgb]{0,0,0}PHOTODIODE}%
}}}}
\put(2941,464){\makebox(0,0)[lb]{\smash{{\SetFigFont{9}{10.8}{\familydefault}{\mddefault}{\updefault}{\color[rgb]{0,0,0}LASER}%
}}}}
\put(1726,539){\makebox(0,0)[lb]{\smash{{\SetFigFont{9}{10.8}{\familydefault}{\mddefault}{\updefault}{\color[rgb]{0,0,0}Y END}%
}}}}
\put(1726,-586){\makebox(0,0)[lb]{\smash{{\SetFigFont{9}{10.8}{\familydefault}{\mddefault}{\updefault}{\color[rgb]{0,0,0}Y MID}%
}}}}
\put(1651,-1036){\makebox(0,0)[lb]{\smash{{\SetFigFont{9}{10.8}{\familydefault}{\mddefault}{\updefault}{\color[rgb]{0,0,0}CORNER}%
}}}}
\put(2701,-1636){\makebox(0,0)[lb]{\smash{{\SetFigFont{9}{10.8}{\familydefault}{\mddefault}{\updefault}{\color[rgb]{0,0,0}X MID}%
}}}}
\put(3751,-1636){\makebox(0,0)[lb]{\smash{{\SetFigFont{9}{10.8}{\familydefault}{\mddefault}{\updefault}{\color[rgb]{0,0,0}X END}%
}}}}
\put(941,492){\makebox(0,0)[lb]{\smash{{\SetFigFont{9}{10.8}{\familydefault}{\mddefault}{\updefault}{\color[rgb]{0,0,0}M}%
}}}}
\put(1201,-598){\makebox(0,0)[lb]{\smash{{\SetFigFont{9}{10.8}{\familydefault}{\mddefault}{\updefault}{\color[rgb]{0,0,0}M}%
}}}}
\put(1531,-1608){\makebox(0,0)[lb]{\smash{{\SetFigFont{9}{10.8}{\familydefault}{\mddefault}{\updefault}{\color[rgb]{0,0,0}M}%
}}}}
\put(2811,-1848){\makebox(0,0)[lb]{\smash{{\SetFigFont{9}{10.8}{\familydefault}{\mddefault}{\updefault}{\color[rgb]{0,0,0}M}%
}}}}
\put(3981,-2288){\makebox(0,0)[lb]{\smash{{\SetFigFont{9}{10.8}{\familydefault}{\mddefault}{\updefault}{\color[rgb]{0,0,0}M}%
}}}}
\put(1301,-1858){\makebox(0,0)[lb]{\smash{{\SetFigFont{9}{10.8}{\familydefault}{\mddefault}{\updefault}{\color[rgb]{0,0,0}M}%
}}}}
\put(1541,-2108){\makebox(0,0)[lb]{\smash{{\SetFigFont{9}{10.8}{\familydefault}{\mddefault}{\updefault}{\color[rgb]{0,0,0}M}%
}}}}
\put(1866,-3983){\makebox(0,0)[lb]{\smash{{\SetFigFont{9}{10.8}{\familydefault}{\mddefault}{\updefault}{\color[rgb]{0,0,0}Y}%
}}}}
\put(1626,-4485){\makebox(0,0)[lb]{\smash{{\SetFigFont{9}{10.8}{\familydefault}{\mddefault}{\updefault}{\color[rgb]{0,0,0}Z}%
}}}}
\put(3695,-5129){\rotatebox{162.0}{\makebox(0,0)[lb]{\smash{{\SetFigFont{9}{10.8}{\rmdefault}{\mddefault}{\updefault}{\color[rgb]{0,0,0}N}%
}}}}}
\put(2181,-4260){\makebox(0,0)[lb]{\smash{{\SetFigFont{9}{10.8}{\familydefault}{\mddefault}{\updefault}{\color[rgb]{0,0,0}X}%
}}}}
\end{picture}%

%% file: blrmsdataflow_testmod.pstex_t
\begin{picture}(0,0)%
\includegraphics{blrmsdataflow_testmod.pstex}%
\end{picture}%
\setlength{\unitlength}{3947sp}%
\begingroup\makeatletter\ifx\SetFigFont\undefined%
\gdef\SetFigFont#1#2#3#4#5{%
  \reset@font\fontsize{#1}{#2pt}%
  \fontfamily{#3}\fontseries{#4}\fontshape{#5}%
  \selectfont}%
\fi\endgroup%
\begin{picture}(3324,4984)(139,-4283)
\put(1651,-1936){\makebox(0,0)[lb]{\smash{{\SetFigFont{9}{10.8}{\familydefault}{\mddefault}{\updefault}{\color[rgb]{0,0,0}ICS-110B A.D.C.}%
}}}}
\put(2551,-1291){\makebox(0,0)[lb]{\smash{{\SetFigFont{9}{10.8}{\familydefault}{\mddefault}{\updefault}{\color[rgb]{0,0,0}16-bit A/D}%
}}}}
\put(1697,-1456){\makebox(0,0)[lb]{\smash{{\SetFigFont{9}{10.8}{\familydefault}{\mddefault}{\updefault}{\color[rgb]{0,0,0}$\times10$}%
}}}}
\put(1126,539){\makebox(0,0)[lb]{\smash{{\SetFigFont{9}{10.8}{\familydefault}{\mddefault}{\updefault}{\color[rgb]{0,0,0}$x$,$y$, or $z$ component}%
}}}}
\put(1126,389){\makebox(0,0)[lb]{\smash{{\SetFigFont{9}{10.8}{\familydefault}{\mddefault}{\updefault}{\color[rgb]{0,0,0}of ground velocity}%
}}}}
\put(376,-361){\makebox(0,0)[lb]{\smash{{\SetFigFont{9}{10.8}{\familydefault}{\mddefault}{\updefault}{\color[rgb]{0,0,0}Guralp CMG-40T seismometer, $H(s)=$}%
}}}}
\put(181,-1291){\makebox(0,0)[lb]{\smash{{\SetFigFont{9}{10.8}{\familydefault}{\mddefault}{\updefault}{\color[rgb]{0,0,0}antialiasing filter}%
}}}}
\put(451,-586){\makebox(0,0)[lb]{\smash{{\SetFigFont{10}{12.0}{\rmdefault}{\mddefault}{\updefault}{\color[rgb]{0,0,0}$\frac{\mathrm{800V/ms^{-1}}(-0.314s^2)(s-159\mathrm{rad/sec})}{(s+150\mathrm{rad/sec})\left|(s-0.02356\mathrm{rad/sec}(-1+i))\right|^2}$}%
}}}}
\put(2679,-1463){\makebox(0,0)[lb]{\smash{{\SetFigFont{9}{10.8}{\familydefault}{\mddefault}{\updefault}{\color[rgb]{0,0,0}$16384$}%
}}}}
\put(2574,-1636){\makebox(0,0)[lb]{\smash{{\SetFigFont{9}{10.8}{\familydefault}{\mddefault}{\updefault}{\color[rgb]{0,0,0}counts/volt}%
}}}}
\put(1126,239){\makebox(0,0)[lb]{\smash{{\SetFigFont{9}{10.8}{\familydefault}{\mddefault}{\updefault}{\color[rgb]{0,0,0}($\mathrm{ms^{-1}}$)}%
}}}}
\put(226,-3826){\makebox(0,0)[lb]{\smash{{\SetFigFont{10}{12.0}{\rmdefault}{\mddefault}{\updefault}{\color[rgb]{0,0,0}points over 1 second; accumulate}%
}}}}
\put(226,-3676){\makebox(0,0)[lb]{\smash{{\SetFigFont{9}{10.8}{\familydefault}{\mddefault}{\updefault}{\color[rgb]{0,0,0}Sum squares of filtered data}%
}}}}
\put(226,-3976){\makebox(0,0)[lb]{\smash{{\SetFigFont{10}{12.0}{\rmdefault}{\mddefault}{\updefault}{\color[rgb]{0,0,0}sum of squares over 1 minute,}%
}}}}
\put(226,-4126){\makebox(0,0)[lb]{\smash{{\SetFigFont{10}{12.0}{\rmdefault}{\mddefault}{\updefault}{\color[rgb]{0,0,0}determine RMS over 1 minute.}%
}}}}
\put(1351,-2536){\makebox(0,0)[lb]{\smash{{\SetFigFont{9}{10.8}{\familydefault}{\mddefault}{\updefault}{\color[rgb]{0,0,0}$H_F(z)=g_F\times$}%
}}}}
\put(1351,-2311){\makebox(0,0)[lb]{\smash{{\SetFigFont{9}{10.8}{\familydefault}{\mddefault}{\updefault}{\color[rgb]{0,0,0}Bandpass Filter:}%
}}}}
\put(226,-2536){\makebox(0,0)[lb]{\smash{{\SetFigFont{9}{10.8}{\familydefault}{\mddefault}{\updefault}{\color[rgb]{0,0,0}decimation}%
}}}}
\put(226,-2701){\makebox(0,0)[lb]{\smash{{\SetFigFont{9}{10.8}{\familydefault}{\mddefault}{\updefault}{\color[rgb]{0,0,0}by $2^N$}%
}}}}
\put(151,-886){\makebox(0,0)[lb]{\smash{{\SetFigFont{9}{10.8}{\familydefault}{\mddefault}{\updefault}{\color[rgb]{0,0,0}injection}%
}}}}
\put(151,-1036){\makebox(0,0)[lb]{\smash{{\SetFigFont{9}{10.8}{\rmdefault}{\mddefault}{\updefault}{\color[rgb]{0,0,0}point}%
}}}}
\put(526,-1036){\makebox(0,0)[lb]{\smash{{\SetFigFont{9}{10.8}{\rmdefault}{\mddefault}{\updefault}{\color[rgb]{0,0,0} A}%
}}}}
\put(1351,-2836){\makebox(0,0)[lb]{\smash{{\SetFigFont{9}{10.8}{\familydefault}{\mddefault}{\updefault}{\color[rgb]{0,0,0}$\frac{1+\sum_{n=1}^8b_nz^{-n}}{1+\sum_{m=1}^8a_mz^{-m}}$}%
}}}}
\end{picture}%